\documentclass{ismdproc}

\begin{document}
\title{Fluctuating initial conditions in hydrodynamics for 
 two-particle correlations}


%

%
\author{{\slshape Y. Hama$^1$\footnote{Speaker}, 
  R.P.G. Andrade$^1$, F. Grassi$^1$ and W.-L. Qian$^2$}\\[1ex]
  $^1$Instituto de F\'{\i}sica, Universidade de S\~ao Paulo, 
  Brazil\\
  $^2$Departamento de F\'{\i}sica, Universidade Federal de 
  Ouro Preto, Ouro Preto-MG, Brazil} 

%

\contribID{xy}  
\confID{yz}
\acronym{ISMD2010}
\doi            

\maketitle

\begin{abstract}
  Event-by-event hydrodynamics, with fluctuating initial 
  conditions, has shown to nicely reproduce several features 
  of experimentally observed quantities in high-energy nuclear 
  collisions. Here we discuss how it may help to understand, 
  in a {\em unified} way, the various structures observed in 
  the long-range two-particle correlations, both in 
  nucleus-nucleus and $p-p$ collisions. Suggestions of how 
  experimentally this description could be tested are also 
  discussed. 
\end{abstract}

\section{Introduction}

In hydrodynamic approach of nuclear collisions, it is assumed 
that, after a complex process involving microscopic collisions 
of nuclear constituents, at a certain early instant a hot and 
dense matter is formed, which would be in local thermal 
equilibrium. 
This state is characterized by some {\it initial conditions} 
(IC), usually parametrized as smooth distributions of 
thermodynamic quantities and four-velocity (see, for 
instance,~\cite{Hirano,Nonaka}). 

However, since our systems are small, {\it important 
event-by-event fluctuations} are expected in real collisions. 
Also, if the thermalization is verified at very early time, 
they should be {\it very bumpy}. 
In previous works, we introduced {\it fluctuating IC} in 
hydrodynamics~\cite{fic,spherio}, by using NEXUS event  generator~\cite{nexus}, and showed important effects on 
several observables. 

In this paper, we briefly survey some of the previous 
results~\cite{fic, granular, v2fluct, hbt} and then discuss more recent results on long-range two-particle correlations. 
 

\section{Some consequences of fluctuating initial conditions}

\begin{figure}[ht]
\centerline
{\includegraphics[width=0.40\textwidth]{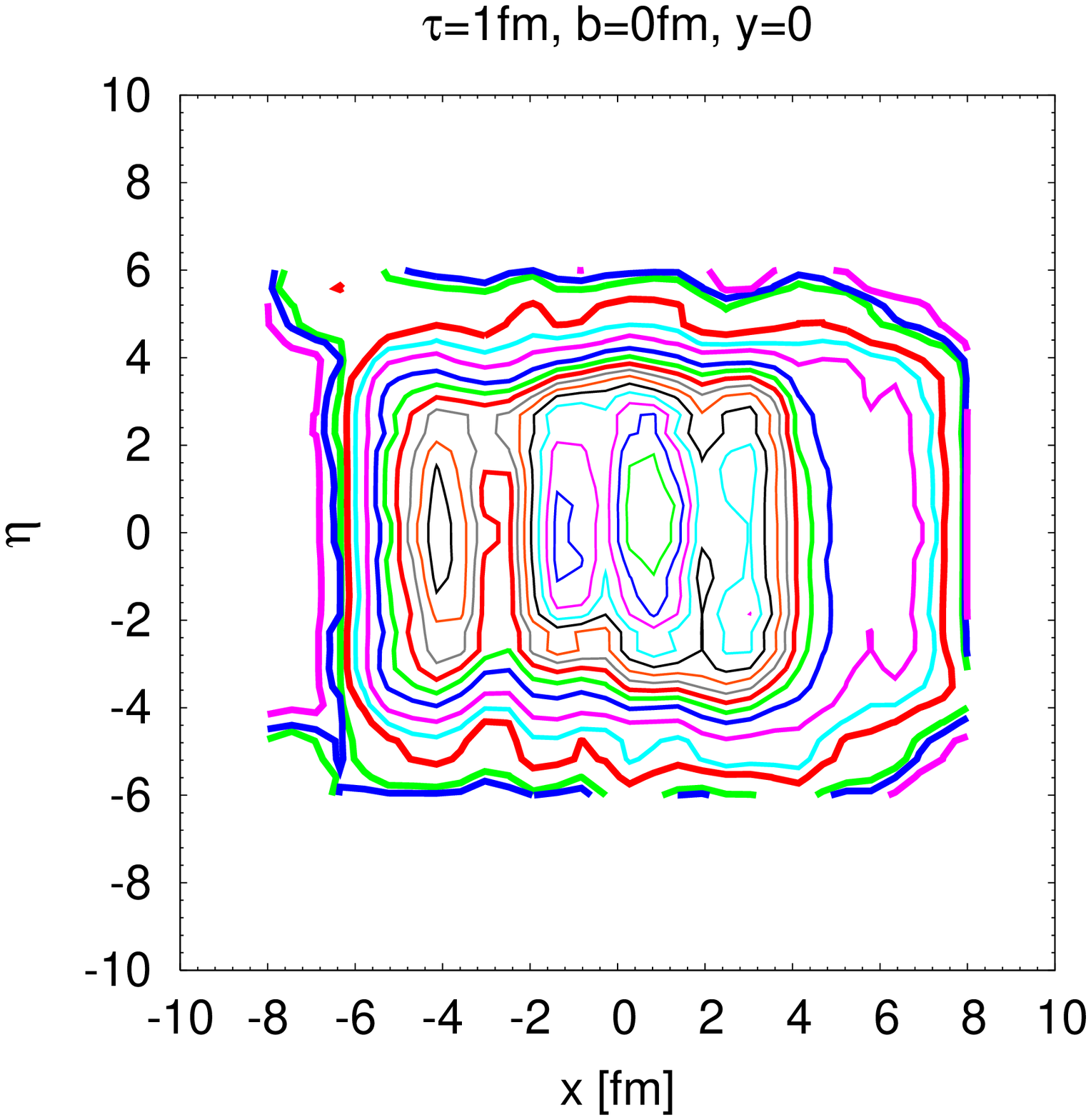}
\hspace{-1.5cm}
\includegraphics[width=0.40\textwidth]{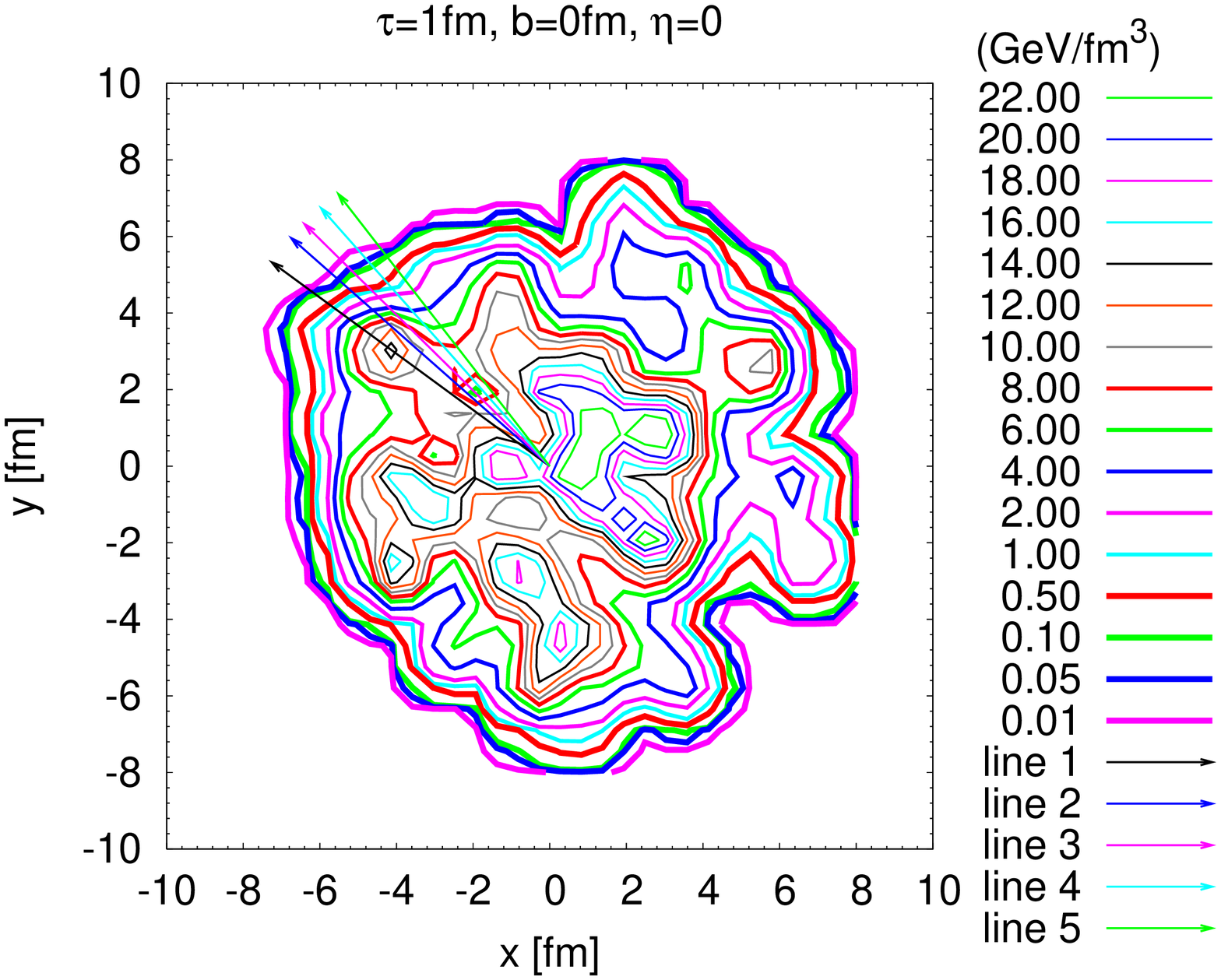}} 
\vspace{-.2cm} 
\caption{NEXUS Fluctuating Initial Conditions.}\label{IC} 
\vspace{-.1cm} 
\end{figure}

In Figure$\,$\ref{IC}, we show the energy-density distribution in a 
typical event of fluctuating IC, generated by NeXuS~\cite{nexus} 
for a central Au+Au collision at 200A GeV. 
Observe that the distribution is very bumpy, as expected in real collisions, having a tubular structure in $\eta$ (leftover from initial particle collisions). 

Some consequences of such high-energy-density spots have been discussed in~\cite{granular}. Because of high concentration of energy in small regions, each tube would suffer a violent explosion and, because of their small size, expand isotropically (in transverse directions). If such a tube is at the surface of the matter, certainly the outgoing part of this initial acceleration would remain, producing high-$p_T$ particles, which would be isotropically distributed in the momentum space. 
Thus, first we expect that high-$p_T$ part of the $p_T$ spectra is enhanced when fluctuating IC are used in our computations, in comparison with the results with averaged (smooth) IC. In the second place, we expect that the elliptic flow coefficient $<v_2>$ suffers reduction as we go to high-$p_T$ region, due to the additional high-$p_T$ isotropic components included now. 
A word of caution is necessary here. We are talking about effects of hot spots and not of the fluctuation, which makes $v_2$ coefficient larger 
in more central windows because it makes the eccentricity bigger~\cite{schenke}. 
As for the $\eta$ dependence of $v_2\,$, we know that the average matter density decreases as $\vert\eta\vert$ increases as reflected in the $\eta$ distribution of charged particles, so when such a blob is formed in the large-$\vert\eta\vert$ regions, its effects appear more enhanced. Therefore, we expect considerable reduction of $v_2\,$ in those regions. All these features have explicitly been verified in~\cite{granular}. 

Another effect of small high-energy-density spots in the IC 
is manifested in the smaller HBT radii, as compared with the case 
of the smooth averaged IC.~\cite{hbt}. This has been shown both by 
using the Cooper-Frye prescription~\cite{cooper-frye}, and by the 
continuous-emission one~\cite{ce}. 

Besides the effects of high-energy-density spots, fluctuations 
of IC imply evidently fluctuations of the resulting observable  quantities. We have discussed that such fluctuations become quite 
large in the anisotropic-flow parameter $v_2\,$~\cite{v2fluct1, v2fluct}, as has been effectively verified by 
experiments~\cite{star_v2f,phobos_v2f}. 

\section{Two-particle correlations in hydrodynamic approach}

One of the most striking results in relativistic heavy-ion 
collisions is the existence of structures in the two-particle 
correlations~\cite{ridge1,ridge1s,ridge2,dhump1,phobos,phoboss} 
plotted as function of the pseudorapidity difference $\Delta\eta$ 
and the angular spacing $\Delta\phi$. The so-called ridge has 
a narrow $\Delta\phi$ located around zero and a long $\Delta\eta$ 
extent. The other structure located opposite to the trigger has 
a single or double hump in $\Delta\phi$; its $\Delta\eta$ extent 
is not well established. More recently, the ridge structure 
has been observed also in $pp$ collisions at LHC~\cite{cms}. 

In an earlier work,~\cite{jun}, we presented evidence that 
hydrodynamic approach reproduces all such structures in heavy-ion 
collisions. In~\cite{jun}, the events computed by using 
the hydrodynamic code SPheRIO~\cite{spherio}, starting from 
event-by-event fluctuating IC, generated by  NeXus~\cite{nexus}, were  analyzed in a similar way to the experimental ones, in particular 
the ZYAM method was used to remove effects of elliptic flow. We later developed a different method to remove elliptic flow from our data 
and checked that all structures are indeed exhibited and other 
features well reproduced (dependence on the trigger- or  associated-particle transverse momentum, centrality, 
in-plane/out-of-plane trigger, etc)~\cite{ismd09, sqm09, bnl10, tube}.

\subsection{Mechanism of ridge formation - one-tube model} 
\label{one-tube}

As seen in Fig.~\ref{IC}, each NEXUS IC is very complicated, so 
difficult to visualize how various structures in the two-particle correlations are generated. In order to clarify the origin of the 
ridge structures, we introduced in \cite{ismd09} a simplified model 
which would allow to follow closely the time development of the fluid 
in the vicinity of one of the high-energy-density tubes. Evidently, 
only those tubes located close to the surface of the hot matter can 
contribute to the correlations. Thus, in our simplified model, 
we replace the complex bulk of the hot matter by the average over 
many events, leaving just one typical tube close to the surface, like 
the one on the {\it line 1} of Fig.~\ref{IC}, right. To simplify the computation, the longitudinal expansion is assumed boost-invariant and 
the transverse expansion is computed numerically (see details in~\cite{ismd09}). 

\begin{figure}[h] 
 \includegraphics[width=4.9cm]{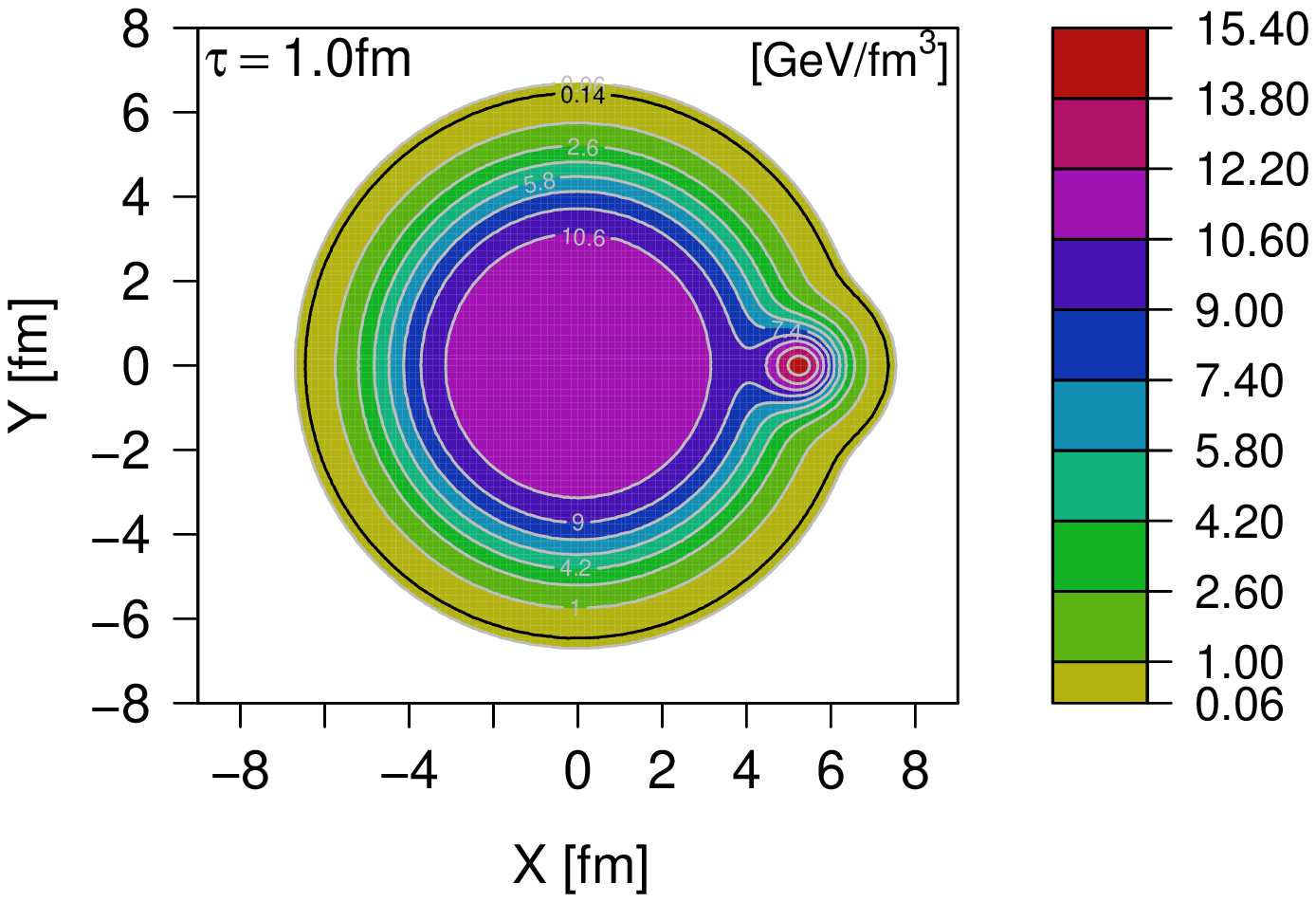}
 \hspace{-.3cm}
 \includegraphics[width=4.9cm]{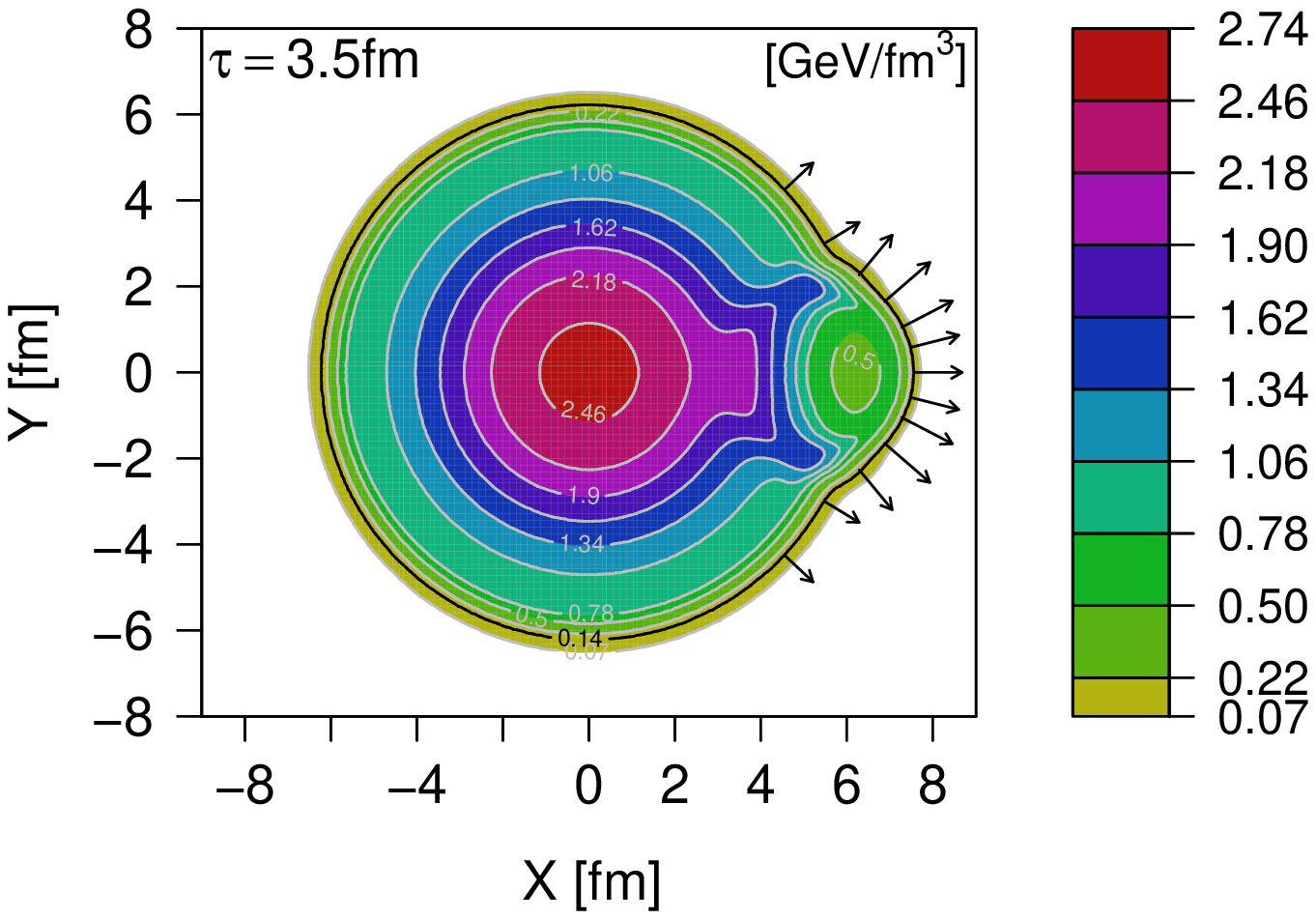}
 \hspace{-.3cm}
 \includegraphics[width=4.9cm]{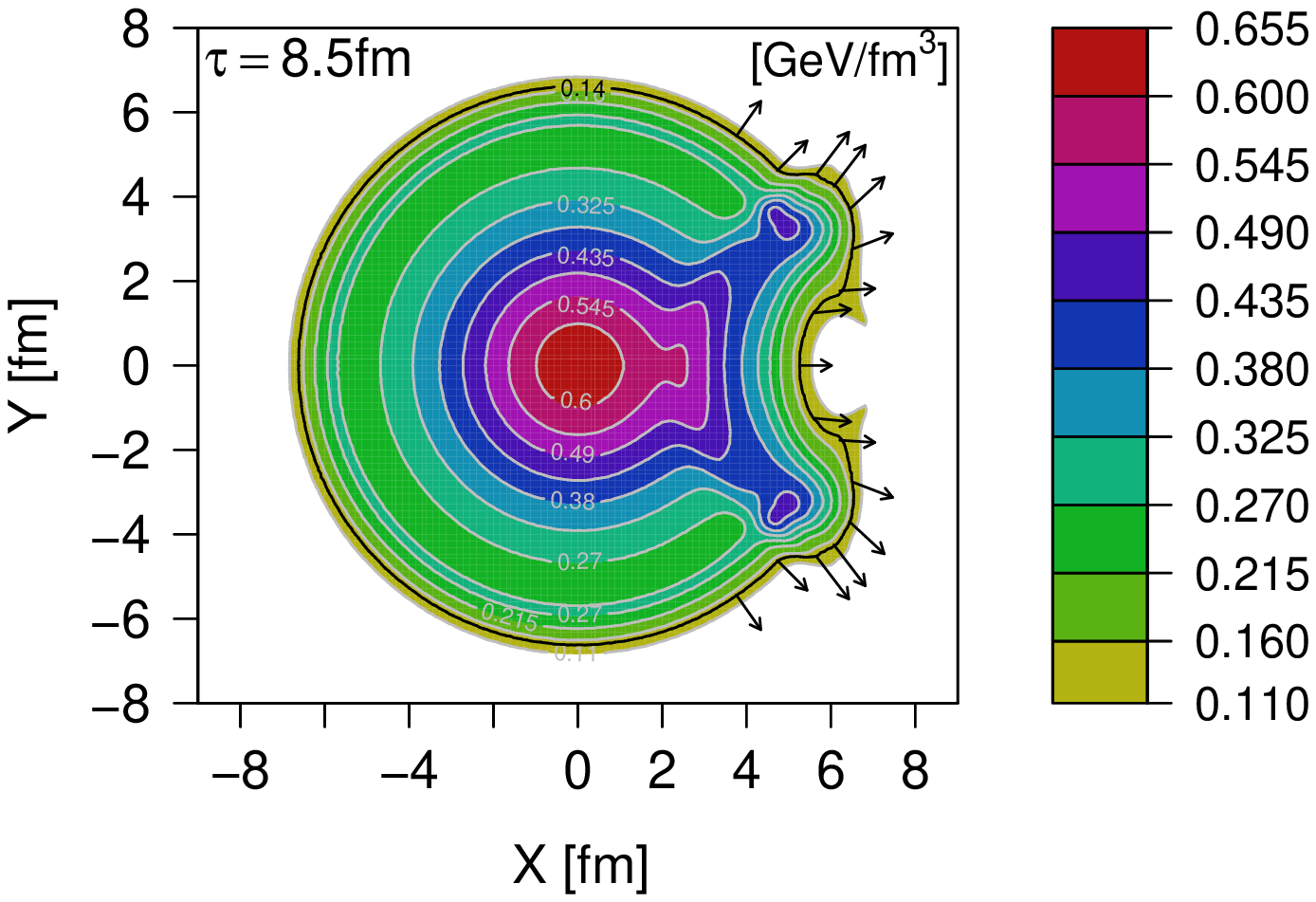}
 \caption{Temporal evolution of energy density 
 for the one-tube model (times: 1.0, 3.5 and 8.5 fm). 
 Arrows indicate fluid velocity on the freeze out surface,  
 thicker curve labeled by the freeze out temperature 0.14
 GeV}\label{flow}  
\end{figure}

Figure~\ref{flow} shows the temporal evolution of the hot matter in this model. As seen, pressed by the violent expansion of the  high-energy-density tube, the otherwise isotropic radial flow of the  background is deflected and guided into two well defined directions, symmetrical with respect to the initial tube position. Notice that the flow is clearly non-radial in these regions.  

\begin{figure}[t] 
\vspace*{-.5cm} 
\centerline
{\includegraphics[width=5.1cm]{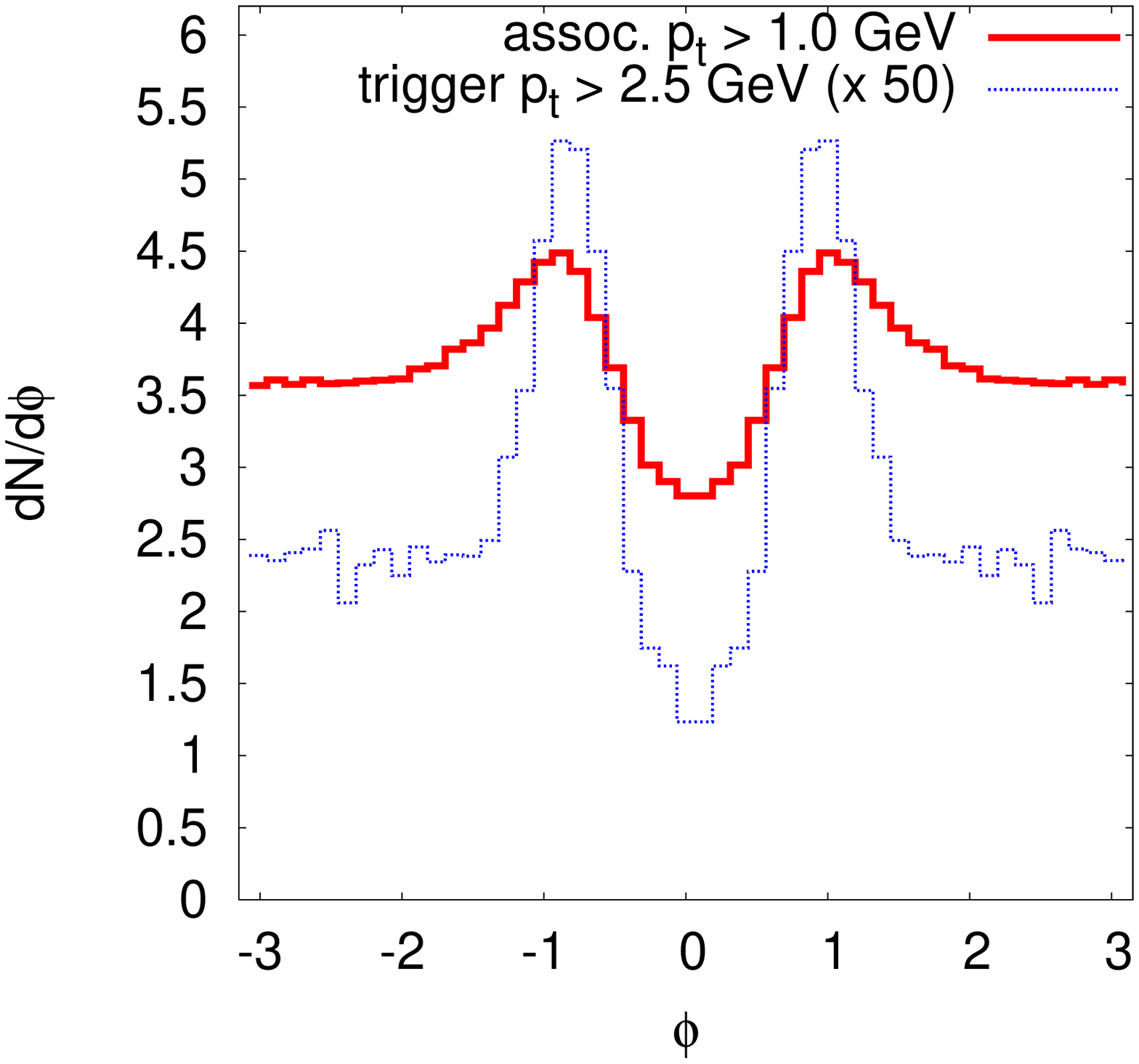} 
\includegraphics[width=5.3cm]{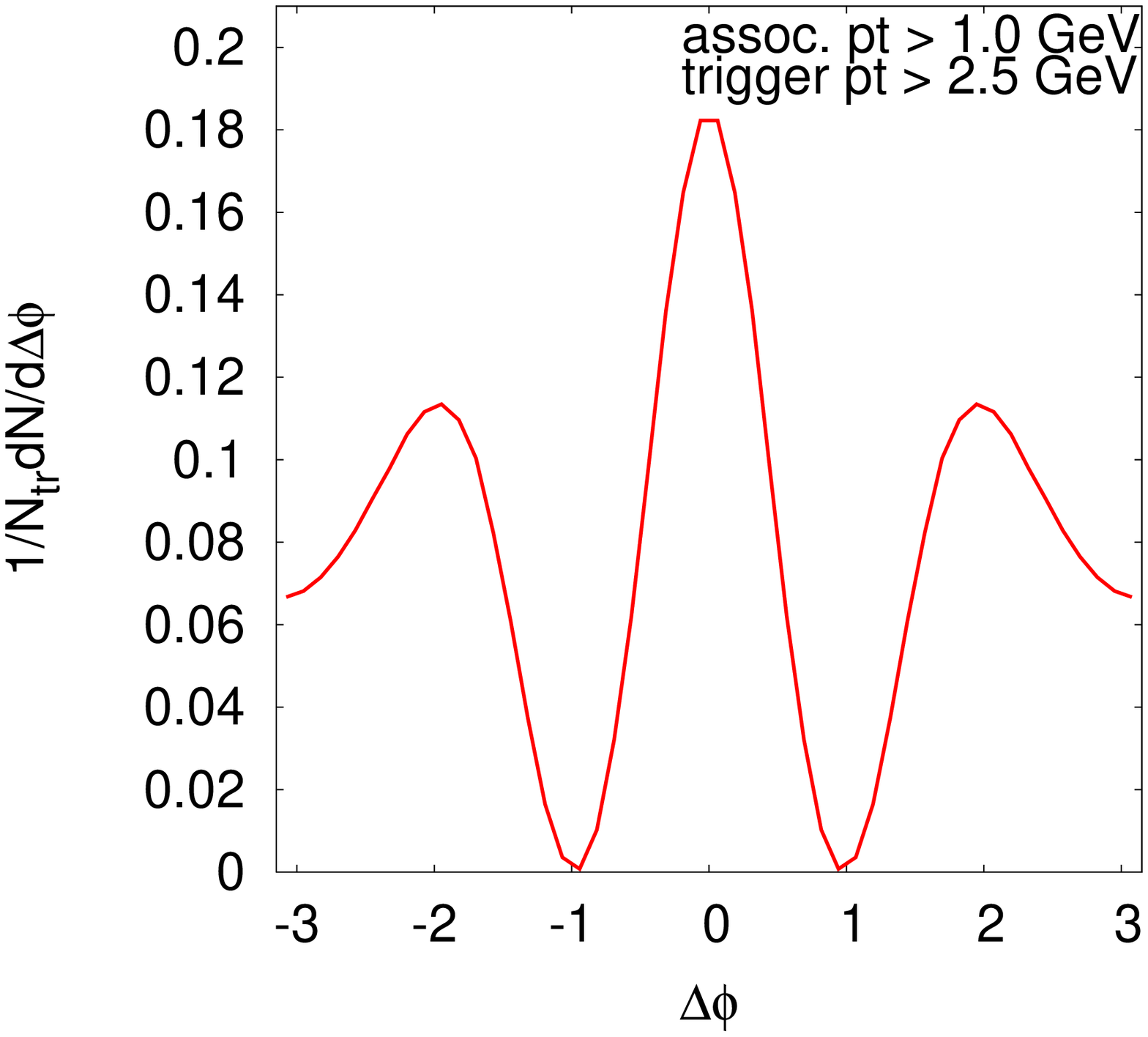}} 
\caption{\label{1dist} 
Angular distributions of particles in some different $p_T$ 
intervals (left) and resulting two-particle correlations (right), in the one-tube model.} 
\end{figure}

The resultant single-particle angular distributions for two different 
$p_T$-intervals are plotted in Fig.~\ref{1dist}, left. As expected they show symmetrical two-peak structures. 
From this plot, we can easily guess how the two-particle angular 
correlation will be. The trigger particle is more likely to be in one 
of the two peaks. We first choose the left-hand side peak. The 
associated particle is more likely to be also in this peak i.e. with 
$\Delta\phi=0$ or in the right-hand side peak with $\Delta\phi\sim+2$. 
If we choose the trigger particle in the right-hand side peak, the 
associated particle is more likely to be also in this peak i.e. with 
$\Delta \phi=0$ or in the left-hand side peak with $\Delta \phi\sim-2$. 
So the final two particle angular correlation must have a large central 
peak at $\Delta\phi=0$ and two smaller peaks respectively at $\Delta\phi\sim\pm 2$. Figure~\ref{1dist} (right) shows that this is indeed the case. 
The peak at $\Delta\phi=0$ corresponds to the near-side ridge and the peaks at $\Delta\phi\sim\pm 2$ form the double-hump ridge. We have checked that this structure is robust by studying the effect of the height and shape of the background, initial velocity, height, radius 
and location of the tube~\cite{ismd09}. 

As stressed above, this simplified description, namely {\it one-tube  boost-invariant model}, has been introduced just to clarify the mechanism of ridge-structure formation. However, it is remarkable that this simple model can describe well so many characteristics observed 
in experiments. 
For a more realistic simulation, we should consider more complex events 
such as NeXus events and average over the fluctuations. For an event   like the one shown in Fig.~\ref{IC}, only the outer tubes need to be  considered. 
The shape of the two-particle correlations for a single tube  
(in particular the peak spacing) is relatively independent
of its features so the various tubes will contribute with 
rather similar two-peaks emission pattern at various angles 
in the single-particle angular distribution. For this single 
event, the two-particle correlation has a well-defined main 
structure similar to that of a single tube (Fig. \ref{1dist}) 
surrounded by several other peaks and depressions due to 
trigger and associated particles coming from different tubes. 
These additional peaks and depressions have positions depending 
on the angle of the tubes between them. When averaging over many randomly fluctuating events these interference terms disappear and 
only the main one-tube like structure is left. 
The main advantage of this interpretation of ridge structures is 
that it involves essentially only the surface of the hot matter. 
The complexity of the kernel does not influence. 

\subsection{Ridge in $pp$ collisions} 

Durig this Symposium, new results on two-particle correlations in 
$pp$ collisions at LHC have been reported~\cite{janssen,cms}. 
Accordingly, high-multiplicity $pp$ collisions show long-range  ($\Delta\eta>4$) nearside ($\Delta\phi\sim0$) ridge structure, which appears more enhanced when intermediate $p_T$ ($1<p_T<3\,$GeV) are considered. 

Such a structure can be produced also within our tube-based description. 
However, since NeXuS generator~\cite{nexus} is based on nucleon-nucleon collisions as an ingredient, 
it can produce only one tube in the IC for $pp$ collisions themselves. 
Evidently, this will not produce ridges. 
%
Some parton based generator is needed to a more complete description of $pp$ data. 

\begin{figure}[h] 
 \includegraphics[width=5.4cm]{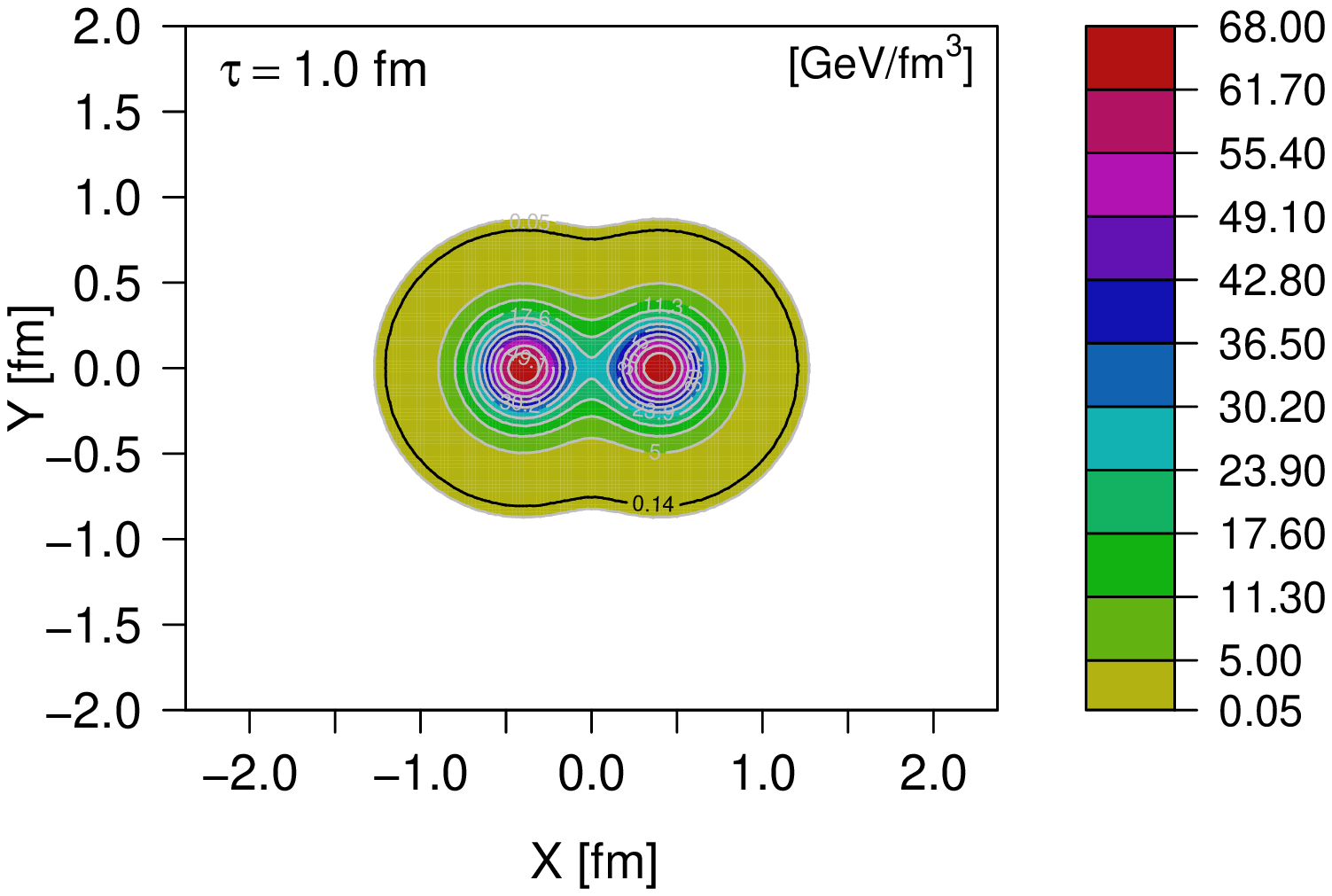}
 \hspace{-.2cm}
 \includegraphics[width=4.8cm]{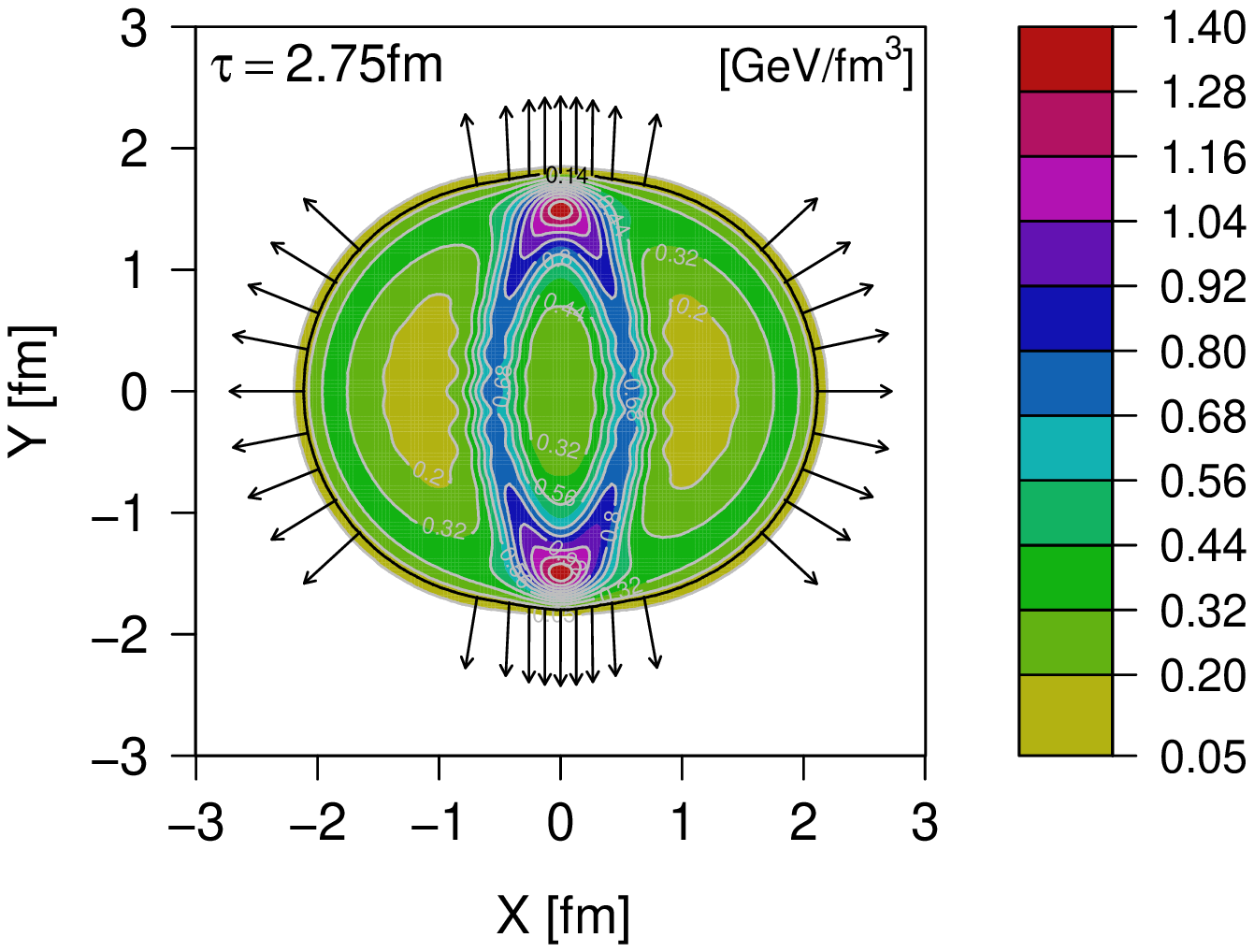}
 \hspace{-.2cm} 
 \vspace*{-1.cm} 
 \includegraphics[width=4.cm]{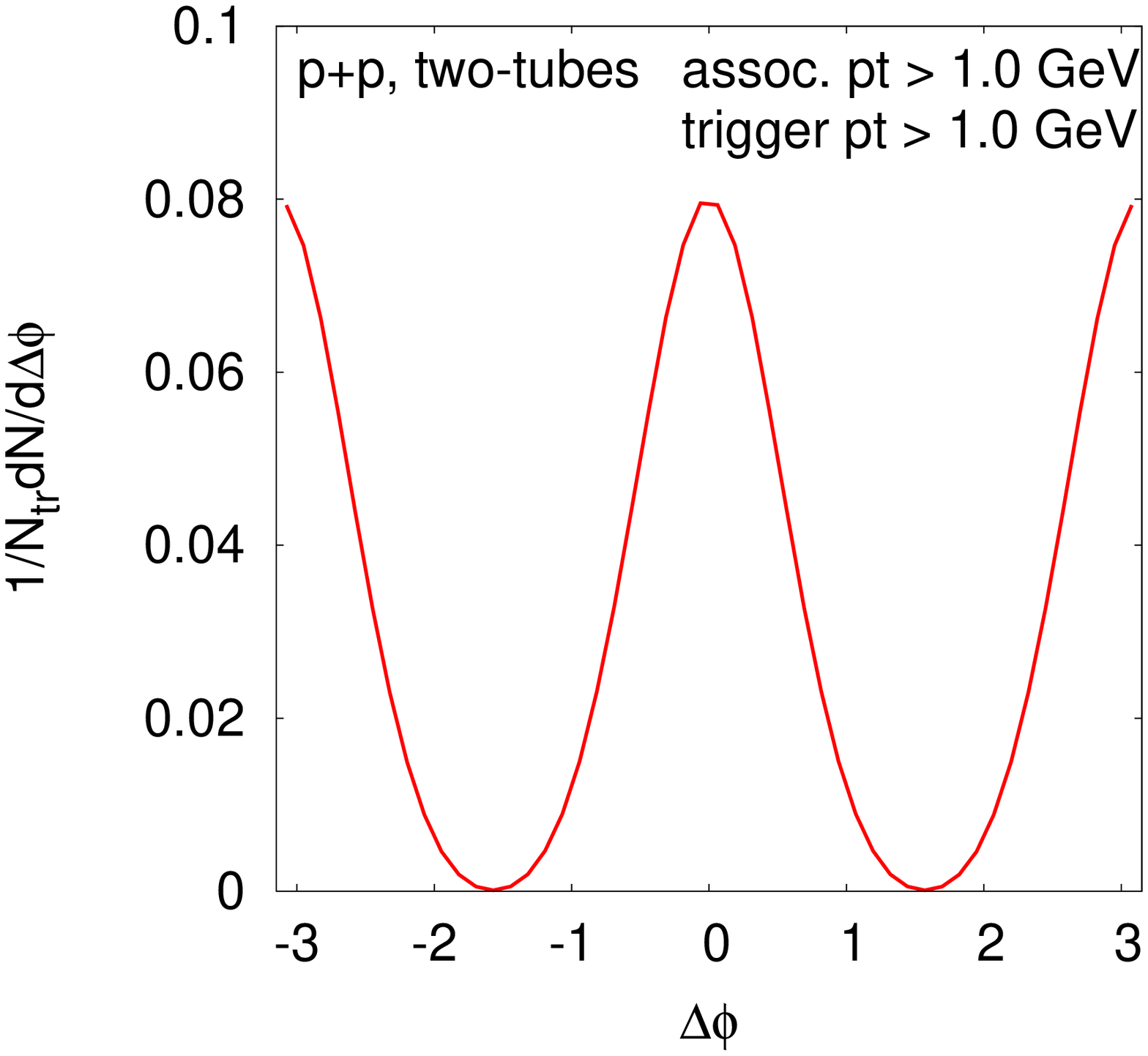} 
 \vspace{1.cm} 
 \caption{Time evolution of two-tube IC for $pp$ at two different 
 instants ($\tau=1$ and $2.75\,$fm), and the resultant two-particle 
 correlation as function of $\Delta\phi$. Here, the normalization has 
 been done by using only one event, so the result should not be compared  
 directly with data.} 
 \label{pp} 
\end{figure}

Here, we shall show how such a structure can be produced within our description, by considering a typical IC in $pp$, formed just by two  tubes. Because of their small size, probably one-tube IC constitute the bulk of the fluctuating IC in high-energy $pp$ collisions. However, although in a small fraction, there should be also events with two tubes produced in the initial collisions of constituents. In Fig.~\ref{pp}, left, we show such two-tube IC. 
K. Werner {\it et al.} discuss nearside ridge, by using EPOS 
generator.~\cite{werner}, where the typical IC shown has a more  elliptical shape or two tubes with smaller distance. 
It is easy to understand that the two tubes begin expanding radially, collide with each other, driving the flow into two opposite directions, as shown in Fig.~\ref{pp}, middle, producing an azimuthal distribution of  particles with two peaks separated by $\pi$. This flow is similar to the  elliptic flow, but probably with sharper peaks. 
Final results of two-particle correlations in such events are two ridges, one at $\Delta\phi\simeq0$ and the other in the awayside at $\Delta\phi\simeq\pi$. In the CMS paper~\cite{cms}, the awayside ridge is not mentioned. In our opinion, this occurred because it is really not visible due to the dominance of jet originated correlations. It would be nice to experimentally separate the ridge from the jet events, by measuring the three-particle correlations in the following way: 

Fix the first associated particle close to the trigger  $(\Delta\phi_1\equiv\phi_1-\phi_t\sim
\Delta\eta_1\equiv\eta_1-\eta_t\sim0)$. Presumably, we are picking jets in this case. On the other side, if we fix $\Delta\phi_1\sim0$ and  $\Delta\eta_1\gtrsim2$, we are 
selecting ridge particles. Then, measure the correlation with a second associated particle in the $(\Delta\phi_2\equiv\phi_2-\phi_t,\Delta\eta_2\equiv\eta_2-\eta_t)$ 
plane. We think that in the first case, 
we will see only typical jet correlations, with a high peak at 
$\Delta\eta\sim\Delta\phi\sim0$ and a high associated ridge at $\Delta\phi_2\sim\pi$. And in the second case two low ridges separated by $\Delta\phi\sim\pi$, eventually with some contaminations of jet events. We call this kind of three-particle correlation {\it 2+1 correlation}~\cite{sqm09,tube}.

\section{Conclusions} 

Hydrodynamic approach starting from event-by-event fluctuating initial conditions, with high-energy-density tubes, has shown 
to reproduce several features of experimentally observed  quantities. 

With regard to the long-range two-particle correlations, it gives a {\em unified} picture for the nearside and awayside structures as observed experimentally. 
In heavy-ion collisions, a high-density tube located close to 
the surface of the hot matter divides the flow coming from inside 
into two currents, producing two-peak angular distribution. This  two-peak distribution is the origin of both the nearside and the  awayside ridges. In high-multiplicity $pp$ collisions, probably 
there are a fraction of events with two high-energy tubes. Such IC produce azimuthal distributions with two peaks separated by $\pi$, resulting in a two-particle correlation with two opposite ridges, one observed as the nearside ridge and the other hindered below the awayside structure associated with jets. 

See contribution by R. Andrade at this Symposium~\cite{rone} 
for detailed discussion of the in-plane/out-of-plane effect within one-tube model. 

\section{Acknowledgments}

We acknowledge funding from FAPESP and CNPq.


\begin{footnotesize}

\end{footnotesize}


\end{document}